\documentclass[12pt]{JHEP3}
\pdfoutput=1

\usepackage{epsfig,cite,multirow} 

\usepackage{amscd}
\usepackage{amsmath,graphicx}
\usepackage{xypic}
\usepackage[matrix,arrow,curve]{xy}
\usepackage{graphicx}
\usepackage{verbatim}
\usepackage{epsf}
\usepackage{latexsym}
\usepackage{amsmath,amsfonts,amssymb,amsthm}
\usepackage{amsmath,amsthm}

\newcommand {\non}{\nonumber}
\newcommand{\Tr}{{\rm Tr}}
\newcommand{\Db}{\Bar{D}}
\renewcommand{\a}{\alpha}
\renewcommand{\b}{\beta}
\renewcommand{\l}{\lambda}
\newcommand{\intsup}{\int\!\! d^3xd^4\theta ~}  
\newcommand{\intk}[1]{\int\!\! \frac{d^3{#1}}{(2\pi)^3} ~}
\newcommand{\intkk}[2]{\int\!\! \frac{d^3{#1}}{(2\pi)^3} \frac{d^3{#2}}{(2\pi)^3} ~}

\title{\begin{center} 
Z-extremization and F-theorem \\  in Chern-Simons matter theories
\end{center}}
\author{
Antonio Amariti$^{1,a}$, Massimo Siani$^{2,b}$
\\~
\\
$^1$Department of Physics, University of California\\
San Diego La Jolla, CA 92093-0354, USA
\\~\\
$^2$Instituut voor Theoretische Fysica, Katholieke Universiteit Leuven,\\
Celestijnenlaan 200D B-3001 Leuven, Belgium.
\\~\\
$^a$\email{antonio.amariti@physics.ucsd.edu} \\
$^b$\email{massimo.siani@fys.kuleuven.be}
}

\abstract{The three dimensional exact $R$ symmetry of $\mathcal{N}=2$ SCFTs  
extremizes the partition 
function localized on a three sphere. Here we verify this statement at weak coupling. 
We give a detailed analysis for  two classes of models.
The first one is an $SU(N)_k$ gauge theory at large $k$ with both fundamental and adjoint matter fields, while 
the second is a flavored version of the ABJ theory, where the CS levels are large but they do not necessarily sum up to zero.
We study in both cases superpotential deformations and compute the $R$ charges at different fixed points.
When these fixed points are connected by an RG flow we explicitly verify that the free energy decreases at the endpoints of the flow 
between the fixed points, corroborating the conjecture of an F-theorem in three dimensions.
}

\preprint{UCSD-PTH-11-06}
\begin{document}

\section{Introduction}

The knowledge of the exact $R$ charges, whose conserved current fits in the same supersymmetry multiplet of the
energy-momentum tensor, is a long standing and interesting problem. In general, along a RG flow, the UV $R$ current
mixes with all the other abelian (flavor) global symmetries of the theory. Moreover, some
new accidental abelian symmetry, not present in the UV description, can appear in the IR, thus making it difficult to
determine the low energy exact $R$ symmetry.

In four-dimensional superconformal field theories (SCFTs) the problem was solved via the $a$-maximization procedure \cite{Intriligator:2003jj}.
It was shown that the exact $R$ symmetry is the most general linear combination of all the global abelian symmetries
which locally maximizes the coefficient $a$ of the conformal anomaly. The latter result is based on the fact that the
coefficient $a$ is
non-perturbatively known \cite{Anselmi:1997am}. A similar argument does not seem to apply to three-dimensional field
theories, for which there is no anomaly. 

An early possibility was discussed in \cite{Barnes:2005bm}. It is based on the observation that 
in $d$ dimensional SCFTs the exact $R$ charge minimizes the coefficient of the $R$-current two point function.
The problem is that in three dimensional theories this quantity receives quantum corrections and a
perturbative analysis is necessary.
More recently it was conjectured that the exact $R$ charge of
$\mathcal{N}=2$ gauge theories in three dimensions with a Chern-Simons term extremizes the absolute value of the partition function
${\cal Z}$ localized on the round sphere $S^3$ \cite{Jafferis:2010un}.
 The path integral becomes exact at one loop, and 
it contains all of the information of the mixing between the $R$ current and the other abelian symmetries.

The latter conjecture has been tested in some examples \cite{Martelli:2011qj,Cheon:2011vi,Jafferis:2011zi,
Amariti:2011hw,Niarchos:2011sn,Minwalla:2011ma}, in which the $R$ charge at the fixed point
  can be extracted by other procedures. For instance, in four-dimensional toric quiver gauge theories 
 the AdS/CFT duality maps the  $a$-maximization procedure
to  the volume extremization  
\cite{Martelli:2005tp,Martelli:2006yb,Gauntlett:2006vf}. In some cases, an
explicit map between the $a$ function and the volumes has been found \cite{Butti:2005vn}. A similar result holds in the large $N$
limit of three-dimensional toric quiver gauge theories \cite{Martelli:2011qj,Cheon:2011vi,Jafferis:2011zi}, in which the partition function is mapped, even before the
extremization, to some volumes of the dual geometrical background. Thus, the extremization of the volumes
provides the exact $R$ charge of the Chern-Simons-matter field theory.
It follows that the partition function is not only able to provide the exact $R$ symmetry, but also that it
is always minimized in this class of theories.

The latter result led to conjecture  \cite{Jafferis:2011zi} that,
as in four dimensions \cite{Cardy:1988cwa},  there is an analogous of the two dimensional  c-theorem
\cite{Zamolodchikov:1986gt}, such that the free energy $F=-\log|\mathcal{Z}|$ reduces at the endpoints of an RG
flow, counting the decreasing of the massless degrees of freedom.

An useful laboratory to check the validity of the $\cal Z$-extremization and the $F$-theorem
are  SCFTs at finite gauge group rank $N$ and large CS level $k$.
Indeed in this case the IR theory admits a weakly coupled description and the perturbative 
$R$ charges, in terms of the small 't Hooft couplings, can be computed.
When the matter fields appear in conjugate representations, namely in $R + \overline R$ representations, we can directly infer
the perturbative $R$ charges from the knowledge of the beta function of an associated $\mathcal{N}=3$ model, making the explicit
two-loop computation unnecessary \cite{Gaiotto:2007qi}. Many different representations were studied in this setting, and
the agreement among the $\mathcal{Z}$ extremization procedure and the perturbative results was shown \cite{Amariti:2011hw}. Moreover, it has been observed that in all
the examples considered the partition function is minimized. The last
statement is of particular interest in corroborating the validity of a $F$-theorem, at least in that regime. More recently, some
negative result for this conjecture in the strong coupling regime has been discussed 
in \cite{Niarchos:2011sn}, due to the presence of accidental symmetries in the IR..

In the present paper we study the weak coupling regime of two different classes of theories. 
In both cases  the matter fields do not sit in the $R+\overline R$ representation, and
the $\mathcal{N}=3$ formalism cannot be used. We check the validity of the $\mathcal{Z}$ extremization with an explicit perturbative computation.
The first is a $SU(N)$ three-dimensional gauge theory coupled to $N_f$ fundamental and $M$ adjoint matter fields.
The second class of models are flavored versions \cite{Bianchi:2009rf,Bianchi:2009ja} of the ABJM \cite{Aharony:2008ug} and ABJ \cite{Aharony:2008gk}, 
whose gravity duals were studied in \cite{Hohenegger:2009as,Gaiotto:2009tk,Hikida:2009tp}.
In both cases the matter fields are not in $R+\overline R$ representation, their $R$ charges do not follow from the simple
$\mathcal{N}=3$ formalism, and a direct quantum computation is necessary. Thus, we compute the $R$ charges at leading order from the
 $\mathcal Z$-extremization procedure and show that they exactly match with the two-loop analysis in all the models we consider.
 Because our results cannot be derived by a more general formalism, this provides a nontrivial test that the exact $R$ symmetry locally
 extremizes the partition function on $S^3$.
Finally, we check in the same perturbative regime the validity of the $F$-theorem by studying some superpotential and higgsing  flow. 
\\
\\
The paper is organized as follows. 
In section \ref{sec1} we review some general property of the partition function of three dimensional $\mathcal{N}=2$ 
field theories localized on the three sphere.
In section \ref{sec2} we study the first class of models, namely $SU(N)_k$ gauge theories with adjoint and fundamental fields.
We show that the two loop perturbative results and the $\mathcal{Z}$ extremization procedure give the same answer for the exact $R$ charge at
 the fixed point, and we check the validity of the $F$-theorem for some RG flow.
In section \ref{sec3} we study along the same lines the models with two gauge groups.
Then we conclude and discuss some possible extension in section \ref{sec4}. 
Some useful formulas for the two loop computations are listed in appendix \ref{A}. In appendix \ref{B} we list the series and the integrals 
necessary for the computation of the partition function.

\section{$\mathcal{Z}$ extremization and $F$ theorem} \label{sec1}

In this section we review the setup necessary to obtain the exact $R$ charge from matrix models.

The basic idea is that the partition function of a three dimensional SCFT $\mathcal{N}=2$ SUSY gauge theory localized on $S^3$ 
is a function of the conformal dimension $\Delta$. The latter is related to the $R$ charge by 
\begin{equation}
\Delta = R = \frac{1}{2}+\gamma
\end{equation}
where $\gamma$ is the anomalous dimension, which takes into account the deviation from the classical value $\Delta=1/2$.

In theories with $\mathcal{N}\geq 3$ supersymmetry the $R$ symmetry group is non abelian and this implies $\Delta=1/2$.
In this case the $S^3$ partition function have been 
computed in \cite{Kapustin:2009kz,Drukker:2010nc}. In $\mathcal{N}=2$ theories the $R$ symmetry becomes abelian
and it can mix with other abelian flavor symmetries along the RG flow.  The partition function in this case is a function of the trial $R_t$ symmetry,
$R_t=R_0+ \alpha_i F_i$, where $R_0$ is the UV $R$ symmetry and $F_i$ are the other abelian symmetries of the theory.
The localized partition function for ${\cal N}=2$ field theories has been explicitly computed in \cite{Jafferis:2010un,Hama:2010av}. It is given by the formula
\begin{equation}
\mathcal{Z}   \equiv \int \prod_{i=1}^{N} du_i e^{ \pi i k \Tr_{F} u^2} {\det}_{\text{adj}}\left( 2 \sinh(\pi u)\right) {\det}_R e^{l(1-\Delta+i u)}
\end{equation}
The first part of this formula is common to theories with a higher degree of supersymmetry.
The integration is over the weights $u_i$ of the fundamental representation.
These variables correspond to the scalars $\sigma$ in the vector multiplet.
They transform in the adjoint representation of the gauge group, and this allows
the integrals to be taken over $\mathbb{R}$.
The determinant over the roots $\rho_{ij}(u)=u_i-u_j$ of the adjoint representation is 
\begin{equation}
 {\det}_{\text{adj}}\left( 2 \sinh(\pi u)\right) =
\prod_{i<j} 4  \sinh^2 (\pi \rho_{ij}(u))
\end{equation}
It is the one loop contribution of every vector multiplet to the partition function.
The gaussian measure, $e^{ \pi i k \Tr_{F} u^2} $,  is the contribution of the CS term at level $k$. The trace is taken in the
fundamental representation and it explicitly becomes Tr$_{F} u^2=\sum_i u_i^2$. Yang-Mills (YM) terms do not
contribute to the partition function, because the YM coupling constant $g_{YM}$ is dimensionful.

The last contribution only appears in $\mathcal{N}=2$ theories.
It is the one loop contribution of the matter fields  ${\det}_R e^{l(1-\Delta+i u)}$, where 
the determinant is over the representation $R$.
More explicitly, we write
\begin{equation}
 {\det}_R e^{l(1-\Delta+i u)} = \prod_{R}  e^{l(1-\Delta+i \rho_i(u))} 
\end{equation}
where $\rho$ are the weights of $R$ in terms of the eigenvalues $u_i$.
The function $l(z)$ arises from the regularization of the one loop determinant of the matter fields. This one loop determinant is 
\begin{equation}
\prod_{n=1}^{\infty} \left(\frac{n+1-\Delta+i \rho(u)}{n+1-\Delta-i \rho(u)} \right)^n
\end{equation}
When the IR $R$ symmetry mixes with other flavor symmetries, the classical result $\Delta=1/2$ 
acquires quantum corrections. By using the Zeta functions the regularized function for the one loop determinant is
$e^{l(z)}$ where
\begin{equation}
l(z) = -\frac{i \pi }{12}-z \text{Log}\left(1-e^{2 i \pi  z}\right)+\frac{1}{2} i \left(\pi  z^2+\frac{Li_2\left(e^{2 i \pi  z}\right)}{\pi }\right)
\label{eq:lz}
\end{equation}
and $z=1-\Delta+i \rho_i(u)$.

In \cite{Jafferis:2010un} 
it was assumed  that  the partition function depends holomorphically  on
the combination $\Delta_j-i m_j$, where $m_j$ is a real mass term for the $j$-th chiral multiplet,
whose conformal dimension is $\Delta_j$.  This holomorphy is not manifest in the computation 
but is explained by the holomorphy of the supersymmetry transformation.
From this assumption it follows that 
$\partial_\Delta  \mathcal{Z} \simeq \partial_m \mathcal{Z} $.
The one point function of an operator in a CFT over $S^3$ is  $\frac{1}{ \mathcal{Z}} \partial_m  \mathcal{Z}
 |_{m=0,\Delta=\Delta_{IR}}$,
and it vanishes at the conformal fixed point if parity is preserved. If parity is broken the 1-point function can be proportional to the identity.
The identity is a parity invariant operator, its VEV is real and Im$(1/ \mathcal{Z} \partial_{m_j} \mathcal{Z})$ vanishes. The final
result is that in this case the exact $R$ charge extremizes $| \mathcal{Z}|^2$. 

The partition function extremization procedure has been verified in many examples, both at weak and strong coupling.
In all these cases the exact $R$ charge actually minimizes the partition function. Even if there is no proof 
that the partition function is minimized by the exact $R$ charge there are other hints suggesting that it should be 
the case.
Indeed the free energy $F$, defined by the relation $|\mathcal{Z}|=e^{-F}$, has been shown to be proportional to
a geometrical  function, called $Z$, which has a unique critical point, the exact $R$ charge (the Reeb vector) \cite{Martelli:2005tp}. 
In four dimensions $Z \sim 1/a$, and it is always minimized. Here the same relation holds between $F$ and $Z$,
leading to the expectation that $\mathcal{Z}$ is always extremized. Moreover as observed in \cite{Jafferis:2011zi}
even in four dimensions the central charge $a$ is related to the free energy on the $S^4$.
These analogies lead to formulate the conjecture that $\Delta F= F_{IR}-F_{UV}<0$ for every RG flow.
This $F$-theorem should imply that the free energy localized on $S^3$ takes into account the reduction of the massless degrees of freedom in any three dimensional RG flow.

\section{$SU(N)_k$ YM CS with fundamental and adjoint fields}\label{sec2}

In this section we  consider a $\mathcal{N}=2$ gauge theory with a Chern-Simons term.
The vector multiplet $V$ is in the adjoint representation of the gauge group $SU(N)$ and we couple it to $M$ adjoint superfields $\phi^i$, $i=1,\ldots,M$
and $N_f$ pairs of chiral fields $q^r, \tilde{q}_r$, $r=1,\ldots,N_f$ in the (anti)fundamental representation of the gauge group and of
the flavor group $SU(N_f)$.

\subsection{R charges from two loop in field theory} \label{1}

We start by computing the $R$ charges in perturbation theory. Our action reads
\begin{equation}
{\cal S} = {\cal S}_{\mathrm{CS}} + {\cal S}_{\mathrm{mat}}
  \label{eq:action} 
\end{equation}
\begin{eqnarray}  
{\cal S}_{\mathrm{CS}}
  &=& \frac{k}{4\pi} \int d^3x\,d^4\theta \int_0^1 dt\: \Tr \Big[
  V \Db^\a \left( e^{-t V} D_\a e^{t V} \right) \Big] \\
{\cal S}_{\mathrm{mat}} &=& \int d^3x\,d^4\theta\: \Tr \left( \bar{\phi}_i
  e^V \phi^i e^{-V} \right)
  \non \\
  &+& \intsup \Tr \left( {\bar q}_r e^V q^r +
  \bar{\tilde{q}}^r \tilde q_r e^{-V} \right)
    \label{eq:mat-action}
\end{eqnarray}
The Chern-Simons level $k$ is an integer number and cannot get quantum corrections in the ${\cal N}=2$ case. Because the action (\ref{eq:action}) does not possess
any continuos parameter and a non-renormalization theorem is at work, our model represents a conformal field theory at the full quantum level. Thus, it is the simplest but
still nontrivial arena to explicitly test the ${\cal Z}$ conjecture in the large $k$ limit.

To proceed, we quantize the theory in a manifest ${\cal N}=2$ setup and compute the two-point correlation functions using the superspace techniques. The quantization of
the theory has been carried out in all the details in \cite{Avdeev:1992jt,Bianchi:2009rf} and we refer the reader to those papers for the details.

\begin{figure}
  \center
  \includegraphics[width=1.0\textwidth]{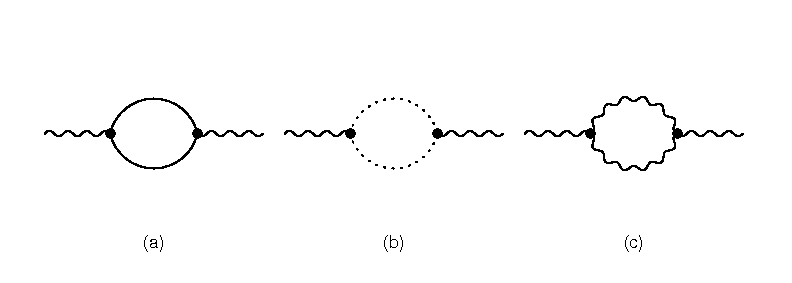}
  \caption{ One--loop diagrams for gauge propagators. }
  \label{fig:gaugeoneloop}
\end{figure}
\begin{figure}
  \center
  \includegraphics{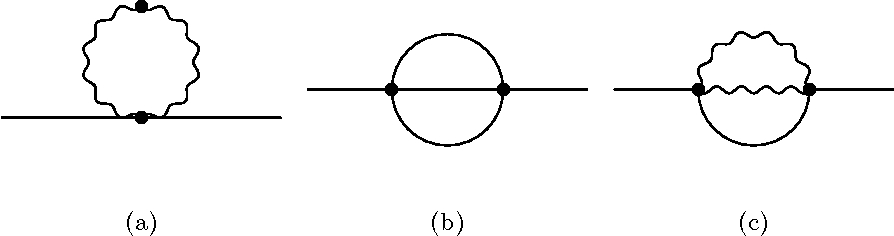}
  \caption{ Two--loop divergent diagrams contributing to the matter propagators.}
  \label{fig:matter2loop}
\end{figure}

At the lowest non-trivial order the $R$ charges can be computed by evaluating the one-loop contribution to the gauge two-point function of figure \ref{fig:gaugeoneloop}
and the two-loop diagrams in figure \ref{fig:matter2loop}.
In the superspace language, we define the superspin-$\frac{1}{2}$ projector
\begin{equation}
  \Pi_{1/2} \equiv \frac{1}{p^2} \Db^\a D^2 \Db_\a (p)
 \label{eq:pi12}
\end{equation}
where $D(p), \Db(p)$ are the covariant derivatives acting on a field of momentum $p$.
In terms of (\ref{eq:pi12}) the finite contributions to the quadratic action for the gauge field are conveniently written as
\begin{equation}
\Pi_{gauge}^{(1)} = \left[ \left(-\frac{1}{8}+ \frac{M}{4} \right) f^{ABC} f^{A^\prime BC} + \frac{N_f}{2} \delta^{AA^\prime} \right] B_0(p) \, p^2 \, V^A(p) \, \Pi_{1/2} \, V^{A^\prime}(-p)
\label{eq:oneloopV}
\end{equation}
where $M$ and $N_f$ are the number of adjoint and fundamental fields respectively, and $B_0(p)=1/(8|p|)$ is the three dimensional bubble scalar integral (\ref{1integral}).
Our conventions for the $SU(N)$ gauge group are given in Appendix \ref{A}.
Like the four-dimensional case, the gauge loop in figure \ref{fig:gaugeoneloop}c cancels against part of the ghost loop of figure \ref{fig:gaugeoneloop}b.
The structure of (\ref{eq:oneloopV}) is a consequence of the supergauge invariance underlying that partial cancellation.

The leading order gauge contributions to the matter fields anomalous dimensions are represented by the diagrams in figure \ref{fig:matter2loop}a and \ref{fig:matter2loop}c,
where we eventually insert (\ref{eq:oneloopV}).
Using the explicit form of the two-loop integral (\ref{integral}), we write the two-loop amplitudes as
\begin{equation}
\begin{split}
\Pi_{\phi}^{(2)} &\equiv \Pi_{\phi} \, \Tr\left( \bar{\phi} \phi \right) \qquad \qquad \Pi_{q}^{(2)} \equiv \Pi_{q} \, \Tr\left( \bar{q} q \right) \\
\Pi_{\phi} &=  - \frac{1}{2 \epsilon} \frac{N}{k^2} \left( M N + N_f + N \right) \\
\Pi_{q} &= - \frac{1}{4 \epsilon} \left( \frac{N^2-1}{N^2 k^2} \right) \left( M N^2 + N N_f -1 \right)
\end{split}
\label{eq:1group}
\end{equation}
where the momentum integrals are performed in $n=3-2\epsilon$ dimensions. The renormalization of the theory now proceeds by defining the renormalized fields
\begin{equation}
 \Phi = Z_\Phi^{-\frac{1}{2}} \, \Phi_B \qquad \bar\Phi = Z_{\bar\Phi}^{-\frac{1}{2}} \, \bar\Phi_B
\end{equation}
together with $k=\mu^{2 \epsilon} \, k_B$ in terms of the renormalization mass $\mu$. Thus, $Z_\Phi \simeq 1-\Pi_\Phi$ and we define the anomalous dimensions as
\begin{equation}
 \gamma_\Phi \equiv \frac{1}{2} \frac{\partial \log{Z_\Phi}}{\partial \log{\mu}} = - \frac{1}{2} \sum_i d_i \nu_i \frac{\partial Z_{\Phi}^{(1)}}{\partial \nu_i}
\label{eq:anodef}
\end{equation}
where $d_i$ is the bare dimension of the $\nu_i$ coupling and $Z_{\Phi}^{(1)} = - \Pi_\Phi$ at the order we are considering.

By direct substitution of (\ref{eq:1group}) into (\ref{eq:anodef}) we find that the $R$ charges in the large $k$ limit are given by
\begin{equation}
\begin{split}
R_\phi &= \frac{1}{2}+\gamma_\phi = \frac{1}{2}-\frac{N(N+N_f+M N)}{k^2} \\
R_q &= \frac{1}{2}+\gamma_q = \frac{1}{2}-\frac{\left(N^2-1\right) (N(N_f+M N)-1)}{2 k^2 N^2}
\end{split}
\label{eq:pert1}
\end{equation}
When $M=0$ we recover the results based on the ${\cal N}=3$ formalism \cite{Gaiotto:2007qi}.

We now consider small deformations of the theory (\ref{eq:action}) by the following superpotential
\begin{equation}
 W = \sum_{i=1}^{M} \alpha_1 \phi_i^4 + \alpha_2 q \phi_i^2 \tilde q + \alpha_3 (q \tilde q)^2
 \label{eq:W1group}
\end{equation}
which represents the most general deformation of our model without changing the field content.
We consider the corresponding diagrams of figure \ref{fig:matter2loop}b and find that
the anomalous dimensions get modified by an amount
\begin{equation}
\begin{split}
 \delta \gamma_q &= \frac{1}{32\pi^2} \left[ \left|\a_2\right|^2  M \frac{N^2-1}{N^2} \left( N^2-2 \right) + 4 \left|\a_3\right|^2 \left(N N_f+1 \right) \right] \\
 \delta \gamma_\phi &= \frac{1}{32\pi^2} \left[ 2 \left|\a_2\right|^2 N_f \frac{N^2-2}{N} + 16 \left|\a_1\right|^2 {\cal J} \right]
\end{split}
\label{eq:deltagamma}
\end{equation}
where
\begin{equation}
 {\cal J} \delta^{AB} \equiv \Tr \left(T^A T^C T^D T^E \right) \, \Tr \left( M T^B T^C \{T^D,T^E\} + T^B T^D \{T^C,T^E\} + T^B T^E \{T^C,T^D\} \right)
\end{equation}

The superpotential (\ref{eq:W1group}) introduces a nontrivial flow from the UV $\a_i=0$ fixed point to a new IR fixed point because all the operators in (\ref{eq:W1group})
are relevant according to (\ref{eq:pert1}). The theory possesses several nontrivial fixed points according to whether some coupling vanishes or not.
 Let us consider the $\a_1 \neq 0$, $\a_3=0$ case first. In the IR, $R_\phi=\frac{1}{2}$ and we distinguish two cases. If the $\a_2$
coupling is also vanishing, the fundamental fields only acquire anomalous dimensions due to their coupling to the gauge fields and their $R$ charge is the same as in
(\ref{eq:pert1}). Otherwise, if they are tree-level coupled to the adjoint fields, at the IR fixed point they are also constrained by $R_q=\frac{1}{2}$
 and the IR theory is finite. A very similar
argument applies whether $\a_3\neq 0$ and $\a_1=0$, with the roles of the fundamental and adjoint fields exchanged. In the $\a_1 \neq 0$, $\a_3 \neq 0$ case, it is easy to
see that the $\a_2$ coupling is an exactly marginal operator, and again $R_q=R_\phi=\frac{1}{2}$ along the whole line of IR fixed points.

We conclude that the most interesting case is the $\a_1=\a_3=0$ one. In that case no subsector of the theory if finite, and we only get the equation $\gamma_q+\gamma_\phi=0$.
From (\ref{eq:deltagamma}) we obtain
\begin{equation}
 \gamma_q\Big|_{\a_1=\a_3=0} = \frac{\left(N^2-1\right) \left(N_f-N_f^2 N+M (M+1) N^3\right)}{k^2 N \left(2 N_f N+M \left(N^2-1\right)\right)} \\
\label{eq:Wpert1}
\end{equation}
and the same result with a minus sign for the adjoints. Note that when $M=0$ we recover (\ref{eq:pert1}) for the fundamental fields, and when $N_f=0$ we have (\ref{eq:pert1})
again for the adjoints, as expected since in those cases no superpotential term is present.

\subsection{R charges from the partition function} \label{2}

The partition function on $S^3$ for our model is given by
\begin{equation}
\mathcal{Z} = 
\int \text{d}u
 f_{vec.}^{(1)} f_{fon.}^{(1)} f_{adj.}^{(1)}
e^{i \pi k Tr(u^2)}
\end{equation}
where
\begin{eqnarray}
f_{vec.}^{(1)} &=&\prod_{i<j} 4\sinh^2(\pi(u_i-u_j))\nonumber \\
f_{fond.}^{(1)} &=&  \prod_{i=1}^{N}\prod_{\eta =\pm1} e^{N_f l(1-\Delta_{q}+i \eta u_i))} \nonumber \\
f_{adj.} &=& \prod_{i<j}\prod_{\eta=\pm 1} e^{M \left( l(1-\Delta_{\phi}+i \eta (u_i-u_j))+ (N-1) l(1-\Delta_\phi)\right)}
\end{eqnarray}
We defined  $\Delta_q$ and  $\Delta_\phi$ respectively as the $R$ charge of the fundamental and of the adjoint fields.

At the perturbative level, large $k$,  $\Delta=1/2+\gamma=1/2+\mathcal{O} (1/k^2)$,
and this allows the expansion of the one loop determinant in $k$ around $\Delta=1/2$.
Moreover in the large $k$ limit the integrals can be evaluated around the stationary point $u_i=0$,
by expanding in $u_i$.
As observed in \cite{Marino:2002fk,Aganagic:2002wv} the expansion of the one loop contribution of the vector field 
factorizes a Vandemronde determinant, simplifying the calculation.
Explicitly we have
\begin{equation} \label{vandfat}
\prod_{i<j}\sinh^2(\pi(u_i-u_j)) = 
\Delta^2(\pi u) e^{{\sum_{p=1}^{\infty}}
\alpha_p \sigma_p^{(g)}}
\end{equation}
where $\alpha_p=\frac{B_{2 p}}{p(2 p)!}$, $B_{2 p}$ are Bernoulli numbers and
\begin{equation}
\sigma_p^{(g)} = \sum_{i<j} (u_i-u_j)^{2 p}
\end{equation}
A similar trick can be  used on the 1-loop matter fields.
In the case of $SU(N)$ gauge groups the traceless condition can be imposed as in \cite{Amariti:2011hw} 
with a $\delta$ function.
By using the Fourier expansion for the $\delta$-function 
\begin{eqnarray}
\delta( \Tr \, x) = \frac{1}{2 \pi} \sum_{m=-\infty}^{+ \infty} e^{i m Tr(x)}
\end{eqnarray}
the trace in the  CS contribution is shifted.
Indeed we have
\begin{equation}
\pi i k_i \Tr u_i^2 + i  m \Tr u_i = -i \frac{N m^2}{4 \pi k } + i \pi k \Tr \left(u+\frac{m}{2 \pi k} \right)^2
\end{equation}
This shift can be absorbed by substituting  the eigenvalues $u_i$ with $u_i-\frac{\beta}{2 \pi}$,
where $\beta\equiv m/k$.
This shift modifies the one loop matter field contribution for the fields in the fundamental 
but not the one for the adjoints.
At large $k$ they become
\begin{eqnarray}
f_{fond.}^{(1)} &=&c_f e^{\sum_{p=1}^{\infty} \xi_p^{(f)} \sigma_{p}^{(f)}}\nonumber \\
f_{adj.}^{(1)} &=&c_g e^{\sum_{p=1}^{\infty}  \xi_p^{(g)} \sigma_{p}^{(G)}}
\end{eqnarray}
where
\begin{eqnarray} \label{espfa}
c_f&=& \frac{2+\gamma_q^2 N_f N}{2^{N_f N+1}}\quad\quad\quad\quad\quad\quad\quad\, \, \,\, , \quad
 \xi_1^{(f)} =N_f C_1(\gamma_q)\quad \!  , \quad 
 \xi_2^{(f)} =N_f C_2(\gamma_q)\nonumber \\
c_g&=& \frac{8+\gamma_{\phi}^2 M(N-1)(N+2)\pi^2}{2^{3+M(N-1)(N+2)/4}}\quad,\quad 
 \xi_1^{(g)} = M C_1(\gamma_\phi)\quad ,\quad
 \xi_2^{(g)} =M C_2(\gamma_\phi) \nonumber 
 \end{eqnarray}
and the functions $C_i(\gamma)$ are given by
\begin{equation}
C_1(\gamma)=-\frac{1}{2}(\pi^2(1+\gamma(\pi^2 \gamma -4)) )\quad,\quad C_2(\gamma) = \frac{1}{12}(\pi^4(1+4 \gamma ( \gamma \pi^2-2)))
\end{equation}
Moreover the exponent of $f_{fond.}^{(1)}$ is
\begin{equation}\
\sigma_p^{(f)} = \sum_{p=1}^{\infty} (u_i-u_\beta)^{2 p}
\end{equation}
where $u_\beta=\beta/(2 \pi)$.
By using these expansions we obtain the 
final expression for the integrand at the two loop order in $1/k$. This expression is 
integrated over the variables $u_i$ and summed over the variables $u_\beta$ 
by using the integrals and the series that
we list in appendix \ref{B}.
After we extremize the resulting partition function we obtain
\begin{eqnarray}
R_\phi &=& \frac{1}{2}-\frac{N(N+N_f+M N)}{k^2} \nonumber \\
R_q &=&  \frac{1}{2}-\frac{\left(N^2-1\right) (N(N_f+M N)-1)}{2 k^2 N^2}
\end{eqnarray}

The matching between the field theory computation and the partition function can be checked even
for other fixed points, when superpotential deformations are added.
For example if we add the small perturbation
\begin{equation} \label{spotRG}
W = \sum_{i=1}^{M} \alpha_1 \phi_i^4 + \alpha_2 q \phi_i^2 \tilde q + \alpha_3 (q \tilde q)^2
\end{equation}
we obtain for the $R$ charges $R=\frac{1}{2} + \gamma$
\begin{center}
\begin{tabular}{c||c|c|c}
& $\alpha_2=\alpha_3=0$&$\alpha_1=\alpha_3=0$&$\alpha_1=\alpha_2=0$\\
\hline
\hline
$\gamma_\phi$ & $0$ & $-\frac{|G|(N_f-N_f^2 N +M(M+1)N^3)}{N(2N_f N+M |G|)k^2}$ & $-\frac{N(N_f+N(M+1))}{k^2}$\\
\hline
$\gamma_q$& $-\frac{|G|(N (N_f+M N)-1) }{2 N^2 k^2}$&$\frac{|G|(N_f-N_f^2 N +M(M+1)N^3)}{N(2N_f N+M |G|)k^2}$ & $0$\\
\end{tabular}
\end{center}
in full agreement with (\ref{eq:pert1}), (\ref{eq:Wpert1}) and discussion above the latter equation.

\subsection{RG flow and $F$-theorem}

In this section we study two different classes of RG flows and verify the validity 
of the $F$-theorem at the endpoints of the flows in both cases.
The first class of RG flow is associated to the higgsing of the gauge symmetry.
We separately study a theory with $N_f$ fundamental fields and a theory with 
$M$ adjoint fields.

In the first case the IR theory has the same field content, with reduced gauge and flavor symmetry,
and some extra singlet.
On the contrary the case with $M$ adjoint fields is more interesting because in
the IR theory there are also new quarks. In this second case we are forced to consider the theory studied above, with both fundamental and adjoint representations.

The second class of RG flows under inspection are the superpotential flows 
with some $\alpha_i \neq 0$ in (\ref{eq:W1group}). In these cases we observe that a four loop computation is necessary, and we restrict to the $G=SU(2)_k$ case.

\subsubsection{Higgsing flow and the $F$-theorem} \label{3}

As usual we higgs the gauge symmetry
by assigning a vev to some field,  introducing a relevant deformation which 
breaks the symmetry and drives the theory to a different IR fixed point.
In general, the vev breaks both the gauge and the flavor symmetries. 
The Goldstone bosons of the
gauge symmetry are eaten by the vector fields which become massive, generating the RG flow. On the other hand, the Goldstone bosons of the flavor symmetry remain
massless uneaten degrees of freedom; depending upon the representation of the field that acquires the vev, they can be either charged or uncharged under the residual gauge symmetry.
For instance, if one of the quarks acquires a vev, say $
\langle q_{N_f,N_c} \rangle  = \langle  \tilde q_{N_c,N_f} \rangle = v
$,
then we are left with $2N_f-1$ massless singlets. In the case that the vev is acquired by a component of the adjoint field, $\langle \phi _{N_c,N_c}\rangle=v$, 
the infrared spectrum contains one singlet, $ N_c-1$ extra charged quarks and $N_c-1$ charged antiquarks. 

In four dimensions the contribution of the eaten matter  lowers both the massless degrees of freedom 
and the central charge $a$, while the uncharged fields work in the opposite direction, rising the value of $a$.
Therefore the
fact that the central charge $a$ decreases at the endpoints of the higgsing flow is a nontrivial check
of the conjectured $a$-theorem, at least in its weak version.
Here we show that some RG flow driven by the Higgs mechanism respects the weakest formulation of the $F$-theorem in three-dimensional Chern-Simons-matter theories at large level $k$.

\subsubsection*{$SU(N)_k$ with fundamentals}

The first model we study has $N_f$ fundamental matter fields $q$ and $\tilde q$ and no superpotential. According to the $D$-term equations we can give
a vev to these fields as before. 
The higgsed theory becomes a $SU(N_c-1)$ gauge theory with $N_f-1$ flavors. There are indeed
$2N_c-1$ components eaten by the vector multiplet.
Moreover are left with $2 N_f-1$ decoupled gauge singlets.
These free fields contribute to the IR theory with an overall contribution.
The free energy difference $\Delta F= F_{IR}-F_{UV}$ at the leading order in the $1/k^2$ expansion is
$$ \Delta F = 
-\frac{1}{2}\log\left(\frac{2^{4 N-3} \,\,k^{2 N-1}  }{\pi ^{2(N-1)}(N-1) \Gamma (N+1)^2}\right)
$$ 
For large $k$ and fixed $N\ll k$ this quantity is negative as expected. It is interesting to observe that 
at the leading order there is no contribution from $N_f$. A possible interpretation is that
if the $F$-theorem holds then the leading effects have to lower the free energy, but as we mentioned above $N_f$ 
works in the opposite direction, because it is associated to the free massless degrees of freedom 
left by the symmetry breaking effects.

In the limit of small $N_f$, as $N$ grows also $k$ has to be kept growing, such that the theory remains perturbative.
In this limit  $\Delta F$  reduces to $-N \log k$.

\subsubsection*{$SU(N)_k$ with one adjoint}

In the theory (\ref{eq:action}) with $M=1$ adjoint field we consider the vev
\begin{equation}
\phi = 
\left(
\begin{array}{cccc}
v&&&\\
&v&&\\
&&\dots &\\
&&&-v \, N 
\end{array}
\right)
\end{equation}
which breaks the gauge symmetry to $SU(N_c-1)$.

In this case the low energy spectrum has only an adjoint charged under the gauge symmetry,
and all the Goldstone bosons get eaten by the vector fields. 
At the leading order the difference $\Delta F$ is 
\begin{equation}\label{leading}
\Delta F =-\frac{1}{2} \log \left(\frac{2^{3 N-2}\,N\,  k^{2 N-1}}{ (N-1) \pi ^{2 N-2} \Gamma (N)^2}\right)
\end{equation}
which is negative for large $k$ and finite $N$.  For large $N$ and $k\gg N$ this expression reduces to
$N \log k$.

\subsubsection*{$SU(N)_k$ with $M$ adjoints}

Tests of the $F$-theorem in the higgsing flow with multiple adjoints are rather nontrivial because in the IR theory there 
are new fundamental fields arising. Indeed when one of the adjoints acquires a vev as above, the other adjoints 
break in one adjoint, one pair of fundamental and antifundamental fields and a singlet.

The IR spectrum of massless degrees of freedom of the corresponding $SU(N-1)_k$ gauge theory is then given by
 $M$ adjoints, $M-1$ pairs of fundamental-antifundamental quarks and $M-1$ singlets.

In this case the theory reduces to the one studied above, with adjoints and fundamentals. Indeed the extra singlets
are completely decoupled free fields, and they add an overall $\exp\left[(M-1) \, l(1/2)\right]$ factor to the partition function, where $l$ is
 defined in (\ref{eq:lz}) and $1/2$ is the $R$ charge of a free field.
At the leading order the difference $\Delta F$ reduces to (\ref{leading}), independently from $M$.
\subsubsection{Superpotential Flow}

Another interesting RG flow is given by the addition of the superpotential (\ref{spotRG}). In this case the computation of the 
two loop partition function is not enough to obtain a non zero result for $\Delta{F}$. Indeed the $\Delta{F}$
function computed at the exact value of the $R$ charge is $\Delta{F}=\mathcal{O}(1/k^4)$. It implies that the knowledge
of the four loop partition function is necessary.

Here we focus on the $G=SU(2)$ case, leaving the answer for generic $SU(N)$ for future works.

We study separately every deformation in (\ref{spotRG}), indeed all of them are relevant at the UV fixed point, $\alpha_i=0$.
We compute the four loop partition function and $R$ charges and substitute back the latter into the former. The resulting
 leading order $\Delta F$s are given by

\begin{center}
\begin{tabular}{c|c}
&$\Delta{F}$\\
\hline
$\alpha_2=\alpha_3=0$&$- \frac{3M(2(M+1)+N_f)^2\pi^2}{k^4}$ \\
$\alpha_1=\alpha_3=0$&$-\frac{3 g N_f(29 + 44 g + 22 N_f)\pi^2 }{256 k^4}$
\\
$\alpha_1=\alpha_2=0$&$-\frac{9N_f(2N_f+4 M-1)^2 \pi^2}{64 k^4}$ 
\\
\end{tabular}
\end{center}
In all these flows the $\Delta F<0$, which confirms the validity of the $F$ theorem in the perturbative window.

\section{Flavored ABJ} \label{sec3}

The second class of models that we study has two gauge groups and it consists of a deformation of 
 the ABJM \cite{Aharony:2008ug} and ABJ \cite{Aharony:2008gk} models. We are interested in adding some fundamental quark and a nontrivial superpotential.
The $\mathcal{N}=3$ flavored theory has been obtained in \cite{Hohenegger:2009as,Gaiotto:2009tk,Hikida:2009tp} and we further generalize the model to
${\cal N}=2$ \cite{Bianchi:2009rf,Bianchi:2009ja} to allow the existence of a nontrivial class of fixed points and RG flows between them.

\begin{figure}
\center
  \includegraphics[width=1.0\textwidth]{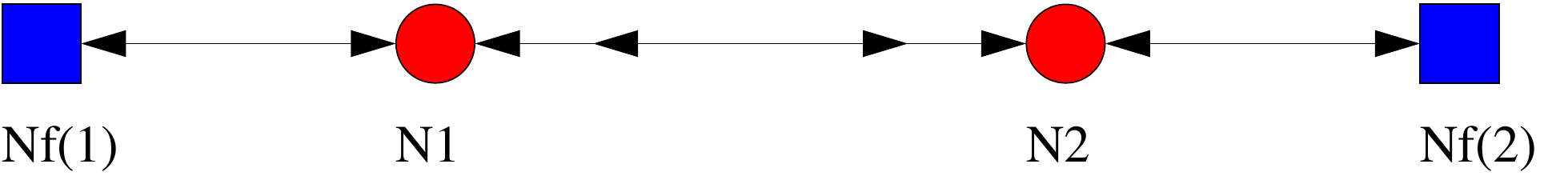}
  \caption{Quiver of the ABJ flavored model  }
  \label{fig:quiverfig}
\end{figure}

The ABJ(M)-like models are specified by two vector multiplets, say $V$ and $\hat V$, in the adjoint representation of the two gauge groups which we will
 take to be $SU(N_1)_{k_1}$ and $SU(N_2)_{k_2}$ respectively. We assign to each vector field the Chern-Simons action (\ref{eq:action}) with two different
 Chern-Simons levels $k_1$ and $k_2$, respectively. We have indicated them by subscribing the
 corresponding gauge group. In addition we have two couples of chiral bifundamental superfields which we indicate as $a_i$ and $b_i$, with $i=1,2$. The
 two $a_i$ transform in the fundamental of the $SU(N_1)_{k_1}$ gauge group and in the antifundamental of $SU(N_2)_{k_2}$. The $b_i$ fields transform in the
 conjugate representation. On top of this, we include fields that are only charged under one of the two gauge groups.
In figure \ref{fig:quiverfig} we give a representation of the resulting theory in terms of a quiver diagram.
The blue boxes represent the flavor $SU(N_f^{(i)})$ symmetries while the red circles are the gauge groups. The lines represents the
fields, where the ingoing  arrows are related to the anti-fundamental representations and the outgoing arrows are related to the 
fundamental ones.
The field content and  the symmetries are specified in the following table.
\begin{center}
\begin{tabular}{|c||cccc|}
\hline
Field &$N_f^{(1)}$&$N_f^{(2)}$&$ N_1$ &$ N_2$\\
\hline
\hline
$a_i$&$1$ &$1$ &$\square$ &$\overline\square$\\
$b_i$&$1$&$1$&$\overline \square$ &$\square$\\
$Q \oplus \tilde Q$&$\square + \overline \square$&$1$&$\overline \square\oplus \square$ &$ 1$\\
$P \oplus \tilde P$&$1$&$\overline \square \oplus \square$&$1$ &$\square\oplus \overline \square$\\
\hline
\end{tabular}
\end{center}
We take the following superpotential
\begin{equation}
W = h_3 \Tr\, a_1 b_1 a_2 b_2 + h_4 \Tr\, a_1 b_2 a_2 b_1 + \alpha_2  \Tr\,  \tilde Q a_2 b_2 Q+
\alpha_3 \Tr\, \tilde P  b_1 a_1 P +  \lambda_3  \Tr\,P \tilde P\, \Tr\,Q \tilde Q
\label{spotABJF}
\end{equation}
where we have explicitly indicated the coupling constants to set the notation for the perturbative results.
The model we choose is, up to equivalent choices of the couplings, the most general one
which allows for superconformal but not finite fixed points which realize the given symmetries \cite{Bianchi:2009rf}.
Moreover, according to whether or not some coupling is set to zero in the UV, the theory presents a nontrivial spectrum of fixed points connected by
 RG flows and exactly marginal operators \cite{Bianchi:2009rf,Bianchi:2009ja}. We will make use of these results both to check the validity of the
 ${\cal Z}$ extremization procedure and to give further evidence of the $F$-theorem.

When all the terms in (\ref{spotABJF}) are kept to be non-vanishing, they and the symmetries of the theory highly constrain the $R$-charges. It is
 easy to see that at the fixed point we have
\begin{eqnarray}\label{Rc}
&&R(a_1)=R(b_1)= R(Q)=R(\tilde Q) =\frac{1}{2} - \gamma\nonumber \\
&&R(a_2)=R(b_2)= R(P)=R(\tilde P) =\frac{1}{2} + \gamma
\end{eqnarray} 
where in the perturbative regime, $k_i \gg N_i,N_f$, the anomalous dimension
$\gamma$ is suppressed by powers of $k_1^n k_2^m$ with $n+m=2$
at the leading order.

We will be also interested in the following two fixed points. The first one is specified by setting $\a_2=\a_3=0$. In this case the anomaly cancellation conditions read
\begin{equation}
\begin{split}
R(a_1)+R(a_2)=R(b_1)+R(b_2)=1 \\
R(P)+R(Q)=R(\tilde P)+R(\tilde Q)=1
\end{split}
\end{equation}
or equivalently
\begin{equation}
\begin{split}
R(a_1)=R(b_1)=\frac{1}{2}+\gamma_1 \qquad
R(a_2)=R(b_2)=\frac{1}{2}-\gamma_1 \\
R(Q)=R(\tilde Q)=\frac{1}{2}+\gamma_Q \qquad
R(P)=R(\tilde P)=\frac{1}{2}-\gamma_Q \\
\end{split}
\end{equation}
The last fixed point we consider is given by the IR theory with $h_i=\l_3=0$ in which
\begin{equation}
\begin{split}
R(a_1)+R(P)=R(b_1)+R(P)=1 \\
R(a_2)+R(Q)=R(b_2)+R(Q)=1
\end{split}
\end{equation}
Note that if one takes $\a_i\neq 0$, $\l_3\neq 0$, then the two operators parametrized by the $h_i$ couplings are exactly marginal. A similar observation
 is that if one takes $h_i \neq 0$ and $\a_i \neq 0$, then the $\l_3$ term is an exactly marginal operator. Thus, in those cases we would get the same
 result as in (\ref{Rc}), as the $R$ charges cannot change by the addition of an exactly marginal operator. The fixed points we are considering exhaust
 a large class of nontrivial tests for the ${\cal Z}$ extremization and the $F$-theorem.

\subsection{Two loop computation}
The superpotential contributions for the theory at hand were already computed in \cite{Bianchi:2009rf}, and they do not change when we consider the $SU(N)$
 case as opposed to the $U(N)$ one. Thus, we only need to compute the two-loop gauge
contributions in the special unitary group case and to add them to the results in \cite{Bianchi:2009rf}.\footnote{Actually, the only computation we
 need is the mixed $(k_1 k_2)^{-1}$ one, as the other one can be inferred from (\ref{eq:pert1}).} The relevant Feynman diagrams are still given in figure
 \ref{fig:gaugeoneloop} and \ref{fig:matter2loop}, and the renormalization procedure is the same outlined in Section \ref{1}. Here, we list the anomalous
 dimensions of all the matter fields in the theory
\begin{equation}
\begin{split}
\gamma_{a_1} &= -\frac{\left(N_2^2-1\right) \left(2 N_1 N_2+N_2 N_f^{(2)}-1\right)}{2 k_2^2 N_2^2}-\frac{\left(N_1^2-1\right) \left(N_1 \left(2 N_2+N_f^{(1)}\right)-1\right)}{2 k_1^2 N_1^2}
\\
&-\frac{\left(N_1^2-1\right) \left(N_2^2-1\right)}{k_1 k_2 N_1 N_2} + \frac{h_3 \bar h_4 + \bar h_3 h_4 +\left(|h_3|^2+|h_4|^2\right) N_1 N_2}{32 \pi^2} + \frac{N_2 N_f^{(2)} |\alpha_3|^2}{32 \pi ^2}
\\
\gamma_{a_2} &= -\frac{\left(N_2^2-1\right) \left(2 N_1 N_2+N_2 N_f^{(2)}-1\right)}{2 k_2^2 N_2^2}-\frac{\left(N_1^2-1\right) \left(N_1 \left(2 N_2+N_f^{(1)}\right)-1\right)}{2 k_1^2 N_1^2}
\\ & - \frac{\left(N_1^2-1\right) \left(N_2^2-1\right)}{k_1 k_2 N_1 N_2} + \frac{h_3 \bar h_4 + \bar h_3 h_4+\left(|h_3|^2+|h_4|^2\right) N_1 N_2}{32 \pi^2} + \frac{N_1 N_f^{(1)} |\alpha_2|^2}{32 \pi ^2}
\\
\gamma_Q &= -\frac{\left(N_1^2-1\right) \left(N_1 \left(2 N_2+N_f^{(1)}\right)-1\right)}{2 k_1^2 N_1^2}+\frac{N_1 N_2 |\alpha_2|^2+N_2 N_f^{(2)} |\lambda_3|^2}{32 \pi ^2} \\
\gamma_P &= -\frac{\left(N_2^2-1\right) \left(N_2 \left(2 N_1+N_f^{(2)}\right)-1\right)}{2 k_2^2 N_2^2}+\frac{N_1 N_2 |\alpha_3|^2+N_1 N_f^{(1)} |\lambda_3|^2}{32 \pi ^2}
\end{split}
\label{eq:abjgamma}
\end{equation}
Note that while the relations (\ref{Rc}) only hold at the fixed point, we find that the two equations $\gamma_{a_i} = \gamma_{b_i}$ always hold
 as a consequence of the residual symmetry of the theory.

We are interested in the values of the $R$ charges at one of the fixed points of the theory. To this end, we find the values of the couplings
which satisfy the fixed point equations $\beta_{\nu_i}=0$ where $\nu_i$ is any coupling of the model and
\begin{eqnarray}
\beta_{h_3} &= 2 h_3 \left( \gamma_{a_1} + \gamma_{a_2} \right) \qquad \qquad \beta_{h_4} = 2 h_4 \left( \gamma_{a_1} + \gamma_{a_2} \right) \\
&\beta_{\lambda_3} =  2 \lambda_3 \left( \gamma_Q + \gamma_P \right) \\
\beta_{\a_2} &= 2 \a_2 \left( \gamma_{a_2} + \gamma_Q \right) \qquad \qquad \beta_{\a_3} = 2 \a_3 \left( \gamma_{a_1} + \gamma_P \right)
\end{eqnarray}
Note that only three of them are independent equations: as discussed above, among the three operators parametrized by $h_3, h_4$ and $\l_3$ there are
 two exactly marginal operators. Once we have solved the fixed point equations, we substitute the fixed points couplings values back into (\ref{eq:abjgamma}).
 The resulting $R$ charges are independent of the coupling constants
\begin{eqnarray}
\gamma_Q = \frac{N_2^2-1}{2 k_2^2 N_2} \frac{N_f^{(2)} \left(N_2\left(2 N_1+ N_f^{(2)}\right)-1\right)}{N_2 N_f^{(2)}+N_1 \left(2 N_2+N_f^{(1)}\right)} -
 \frac{N_1^2-1}{2 k_1^2 N_1} \frac{N_f^{(1)} \left(N_1 \left(2 N_2+N_f^{(1)}\right)-1\right)}{N_2 N_f^{(2)}+N_1 \left(2 N_2+N_f^{(1)}\right)}\nonumber \\
\label{pertmax}
\end{eqnarray}
We started with five different couplings and fixed three of them by the fixed point equations.
The fact that our result does not depend upon the remaining two coupling constants is not surprising, because they only
parametrize exactly marginal operators \cite{Bianchi:2009rf} along which the $R$ charges are constants.

There is an interesting RG flow between the IR $\alpha_i \neq 0$ fixed point of (\ref{eq:abjgamma}) with anomalous dimension given by (\ref{pertmax})
 and another UV fixed point with $\alpha_2=\alpha_3\equiv \alpha=0$. Solving for the latter, we find
\begin{equation}
\begin{split}
\gamma_Q \Big|_{\alpha=0} &= \frac{N_2^2-1}{2 k_2^2 N_2} \frac{N_f^{(2)} \left(N_2\left(2 N_1+ N_f^{(2)}\right)-1\right)}{N_2 N_f^{(2)}+N_1 N_f^{(1)}} -
 \frac{N_1^2-1}{2 k_1^2 N_1} \frac{N_f^{(1)} \left(N_1 \left(2 N_2+N_f^{(1)}\right)-1\right)}{N_2 N_f^{(2)}+N_1 N_f^{(1)}} \\
\gamma_{a_1} \Big|_{\alpha=0} &= 0
\label{pertmaxa0}
\end{split}
\end{equation}
While the absolute value of the anomalous dimensions decreases during the RG flow, it is interesting to see whether the free energy is a monotonic function.

Finally, at the $h_i=\l_3=0$ fixed point we get
\begin{equation}
 \begin{split}
  \gamma_Q\Big|_{h_i=\l_3=0} &= \frac{\left(N_1^2-1\right) \left(N_2^2-1\right)}{k_1 k_2 N_1 \left(N_2+N_f^{(1)}\right)}+\frac{\left(N_2^2-1\right) \left(2 N_1 N_2+N_2 N_f^{(2)}-1\right)}{2 k_2^2 N_2 \left(N_2+N_f^{(1)}\right)} \\
           &+\frac{\left(N_1^2-1\right) \left(N_2-N_f^{(1)}\right) \left(N_1 \left(2 N_2+N_f^{(1)}\right)-1\right)}{2 k_1^2 N_1^2 \left(N_2+N_f^{(1)}\right)} \\
  \gamma_P\Big|_{h_i=\l_3=0} &= \frac{\left(N_1^2-1\right) \left(N_2^2-1\right)}{k_1 k_2 N_2 \left(N_1+N_f^{(2)}\right)}+\frac{\left(N_1^2-1\right) \left(N_1 \left(2 N_2+N_f^{(1)}\right)-1\right)}{2 k_1^2 N_1 \left(N_1+N_f^{(2)}\right)} \\
           &+\frac{\left(N_2^2-1\right) \left(N_1-N_f^{(2)}\right) \left(2 N_1 N_2+N_2 N_f^{(2)}-1\right)}{2 k_2^2 N_2^2 \left(N_1+N_f^{(2)}\right)}
 \end{split}
\end{equation}

\subsection{$\mathcal{Z}$ extremization}
The partition function localized on $S^3$ of the flavored ABJ model is 
\begin{eqnarray}
\mathcal{Z} =  \int du dv f_{vec.}^{(1)}  f_{bif.}^{(1)} f_{fond.}^{(1)} e^{i \pi^2 \left(k_1 \text{Tr} u^2 +k_2\text{Tr} v^2 \right)} 
\delta(\text{Tr u})\delta(\text{Tr v})
\end{eqnarray}
where
\begin{eqnarray}
f_{vec.}^{(1)} &=&\prod_{i<j}4\sinh^2(\pi(u_i-u_j)) \prod_{\tilde i < \tilde j} 4\sinh^2 (\pi(v_{\tilde i}-v_{\tilde j}))\nonumber \\
f_{bif.}^{(1)} &=& \prod_{i=1}^{N_1}\prod_{\tilde j=1}^{N_2}\prod_{\eta,\rho =\pm1} e^{l(1/2+\eta \gamma+\rho i (u_i-v_{\tilde j}))}\\ 
f_{fond.}^{(1)} &=&  \prod_{i=1}^{N_1}\prod_{\tilde j=1}^{N_2}\prod_{\eta,\rho =\pm1}  e^{N_f(l(1/2-\gamma +\rho i u_i)+l(1/2+a +\rho i v_{\tilde j}))} \nonumber
\end{eqnarray}
and the $\delta$-functions enforce the $SU(N_i)$ traceless condition.

In the large $k_i$ limit the contribution of the vector multiplets are the same as (\ref{vandfat}), where a Vandermonde
determinant is factored out and the loop corrections can be computed order by order in $1/k_i$.
The fundamental matter fields contributions are also expanded as in (\ref{espfa}).
Similarly the bifundamental fields can be expanded as
\begin{equation}
f_{bif.}^{(1)} = c_b e^{\sum_{p=1}^{\infty}  \xi_p^{(b)} \sigma_{p}^{(b)}}
\end{equation}
where $c_b=\frac{1+\gamma^2 N_1 N_2 \pi^2}{4^{N_1 N_2}}$ and 
\begin{equation}
\sigma_p = \sum_{i,\tilde j}\left(u_i-v_{\tilde j} - u_{\beta_1} +u_{\beta_2}\right)^{2p}
\end{equation}
The relevant coefficients $\xi_p^{(b)}$ for the two loop computations are
\begin{equation}
\xi_1^{(b)} = -\pi^2(1+\gamma^2 \pi^2) \quad,\quad \xi_2^{(b)} = \frac{\pi^4}{3}\left(\frac{1}{2}+2 \gamma^2 \pi^2 \right)
\end{equation}

By expanding the partition function and by extremizing $|\mathcal{Z}|^2$ to respect to $a$ we obtain
\begin{equation}
\gamma = 
\frac{N_f^{(2)} |G_2| \left(2 N_1 N_2+N_f^{(2)} N_2-1\right)}{2 k_2^2 N_2 \left(2 N_1 N_2+ N_f^{(1)} N_1+N_f^{(2)}N_2\right)}-
\frac{N_f^{(1)} |G_1| \left(2 N_1 N_2+N_f^{(1)} N_1-1\right)}{2 k_1^2 N_1 \left(2 N_1 N_2+ N_f^{(1)}N_1+N_f^{(2)}N_2\right)}
\end{equation}
which matches  with the perturbative result (\ref{pertmax}) computed above.

Even in this case there are other fixed points if some of the superpotential couplings is set to zero.
Indeed we have three independent equations for the beta functions and we can set some of them to zero.
We studied three different situations, $h_3=h_4=0$, $\lambda_3=0$ and $\alpha_1=\alpha_2=0$
We obtained \\
\begin{center}
\begin{tabular}{c|||c} 
$\gamma$&$h_3=h_4=\lambda_3=0$\\
\hline
\hline
$\gamma_{a_1}$ &$-\frac{1}{ N_f^{(1)}+N_2}\left(\frac{|G_1| (N_2-N_f^{(1)})(2 N_1 N_2+N_f^{(1)} N_1-1) }{2 k_1^2 N_1^2}+\frac{ |G_1||G_2| }{ k_1 k_2 N_1}+\frac{|G_2| (2 N_1 N_2+N_f^{(2)} N_2-1)}{2k_2^2   N_2} \right)$  \\
\hline
$\gamma_{a_2}$ & $-\frac{1}{ N_1+N_f^{(2)}}\left(\frac{|G_2| (N_1-N_f^{(2)})(2 N_1 N_2+N_f^{(2)} N_2-1)}{2 k_2^2  N_2^2}+\frac{ |G_1||G_2|} {k_1 k_2  N_2}+\frac{|G_1|(2 N_1 N_2+N_f^{(1)} N_1-1) }{2 k_1^2  N_1} \right)$ \\
\hline
$\gamma_{Q}$ & $~~\frac{1}{ N_f^{(1)}+N_2}\left(\frac{|G_1| (N_2-N_f^{(1)})(2 N_1 N_2+N_f^{(1)} N_1-1) }{2 k_1^2 N_1^2}+\frac{ |G_1||G_2| }{ k_1 k_2 N_1}+\frac{|G_2| (2 N_1 N_2+N_f^{(2)} N_2-1)}{2k_2^2   N_2} \right)$ \\
\hline
$\gamma_{P}$ & $~~\frac{1}{ N_1+N_f^{(2)}}\left(\frac{|G_2| (N_1-N_f^{(2)})(2 N_1 N_2+N_f^{(2)} N_2-1)}{2 k_2^2  N_2^2}+\frac{ |G_1||G_2|} {k_1 k_2  N_2}+\frac{|G_1|(2 N_1 N_2+N_f^{(1)} N_1-1) }{2 k_1^2  N_1} \right)$  \\
\end{tabular}
\end{center}
~
\\
~
\begin{center}
\begin{tabular}{c|||c} 
$\gamma$&$ \alpha_1=\alpha_2=0$\\
\hline
\hline
$\gamma_{a_1}$ &$0$  \\
\hline
$\gamma_{a_2}$ & $0$ \\
\hline
$\gamma_{Q}$ &  $ - \frac{1}{N_f^{(1)} N_1+N_f^{(1)}N_1}
\left(\frac{N_f^{(1)}|G_1|  \left(N_1 \left(2N_2+N_f^{(1)}\right)-1\right) }{2 k_1^2 N_1}
-
\frac{N_f^{(2)}|G_2|  \left(N_2 \left(2N_1+N_f^{(2)}\right)-1\right) }{2 k_2^2 N_2}\right) $\\
\hline
$\gamma_{P}$ &  $~~ \frac{1}{N_f^{(1)} N_1+N_f^{(2)}N_2}
\left(\frac{N_f^{(1)}|G_1|  \left(N_1 \left(2N_2+N_f^{(1)}\right)-1\right) }{2 k_1^2 N_1}
-
\frac{N_f^{(2)}|G_2|  \left(N_2 \left(2N_1+N_f^{(2)}\right)-1\right) }{2 k_2^2 N_2}\right)  $\\\end{tabular}
\end{center}
When $\alpha_i \neq 0 \neq h_i$ the coupling $\lambda_3$ is an exactly marginal deformation at the fixed point and
indeed the anomalous dimensions and the $R$ charge coincide with (\ref{pertmax}) in this case.
Indeed this operator does not break any global symmetry left by the superpotential.

\subsection{$F$-theorem}

In this section we discuss the validity of the conjectured $F$-theorem for the flavored ABJ model.
We study some superpotential flow, interpolating among the fixed points discussed in the last two sections.
As in the one gauge group case we restrict our attention to the $SU(2)_{k_1} \times SU(2)_{k_2}$ theory, with
$N_{f}^{(1)}+N_{f}^{(2)}$ flavors. While we allow the two Chern-Simons levels to acquire different absolute values, we leave the general $SU(N_1) \times SU(N_2)$
 case for future computations.

The first RG flow starts with $\alpha_i=0$ and $h_i\neq0\neq \lambda_3$. When we add the small $\alpha_i$ superpotential deformation we reach the IR fixed point
 specified by (\ref{pertmax}). The constraint $\gamma_P+\gamma_Q=0$ imposes that one of the operators in (\ref{spotABJF})  is
relevant and the other is irrelevant. Anyway the existence of the superconformal IR fixed point 
for this deformed theory justifies the existence of an RG flow \cite{Bianchi:2009rf,Bianchi:2009ja}, and implies that the irrelevant operator is
actually 
 dangerously irrelevant.
 Indeed, supposing that $\gamma_P>0$, then the fields $a_2$ and $b_2$ acquire 
 a positive anomalous dimension. At the same time the operator $W_{abj}$ forces the $R$ charge of
 $a_1$ and $b_1$ to become negative, and this makes the operator 
 $Q a_1 b_1 \tilde Q$ relevant during the flow. Finally, at the fixed point all the operators
 are exactly marginal and we obtain the theory studied above. The same discussion holds if $\gamma_Q>0$.
The free energy difference $\Delta F= \log{\frac{|Z_{UV}|}{|Z_{IR}|}}$ vanishes at the two loop level, while at four loop it is
\begin{equation}
\Delta F = -\frac{9 \pi^2 \left(N_f^{(1)}\left(7+2 N_f^{(1)}\right) k_2^2-N_f^{(2)}\left(7+2 N_f^{(2)}\right) k_1^2\right)^2}
{16 k_1^4 k_2^4 \left(N_f^{(1)}+N_f^{(2)}\right)\left(4+N_f^{(1)}+N_f^{(2)}\right)}
\end{equation}
Note that this quantity is never positive. There are some special values of $N_f^{(i)}$ and $k_i$ at which $\Delta F=0$. 
For example if $|k_1|=|k_2|$, $N_f^{(1)}=N_f^{(2)}$, $\Delta F=0 $ but this is only a consequence of the fact that the global symmetry is enhanced, and this makes
 the operators associated to $\alpha_i$ exactly marginal deformations. Thus, 
a vanishing free energy difference would have to be expected in this case.
Moreover, other values of $N_f^{(i)}$ and $k_i$ can lead to $\Delta F=0 + \mathcal{O}(1/k^6)$.
In those cases the validity of the $F$-theorem 
has to be checked at six loop.

The second RG flow starts from the theory with $h_i=0$ and $\lambda_3=0$ in the UV. We then add these deformations and flow to the 
same IR theory above, where (\ref{pertmax}) holds. In this case $\Delta F$ is given by
\footnotesize
\begin{eqnarray}
&&\Delta{F} =\\
 &&-\frac{9 \pi^2\!\!\left(\! \left(\!N_f^{(1)}\!\left(\!N_f^{(2)}\!\!\!-\!\!2\right)\!-\!\!8\!\right)\left(\!2N_f^{(2)}\!+\!7\right)\! k_1^2\!
+\!\! \left(\!N_f^{(2)}\left(\!N_f^{(1)}\!\!\!-\!\!2\right)\!-\!\!8\!\right)\!\left(\!2N_f^{(1)}\!+\!7\right)\! k_2^2\! \!-\!\!12\!\left(\!N_f^{(1)}\!\!+\!\!N_f^{(2)}\!+\!4\!\right)\!k_1 k_2\!\right)^2}{64k_1^4k_2^4\left(N_f^{(2)}+2\right)\left(N_f^{(1)}+2\right)\left(N_f^{(1)}+N_f^{(2)}+4\right)}\nonumber
\end{eqnarray}
\normalsize
Even in this case the difference is never positive. For values of $N_f^{(i)}$ and $k_i$ such that $\Delta F=0$ a six loop computation is necessary.

\section{Conclusions and discussion} \label{sec4}

We computed the exact $R$ charges for two wide classes of $\mathcal{N}=2$ three-dimensional Chern-Simons-Matter theories at large level, both via the recently
 proposed $\cal Z$ extremization method and by carrying out the explicit weakly coupled computations. The two ways have been found to agree at the order we
 considered, thus providing another evidence for the conjecture of the $\cal Z$ extremization method.
We first considered the class of $SU(N)_k$ gauge theories coupled to an arbitrary number of flavor and adjoint fields. 
We showed that the two loop side of the computation and the $\cal Z$ extremization match even in the presence of a superpotential,
where the theory possesses a nontrivial spectrum of fixed points. 
We verified with an analytical computation the validity of the $F$-theorem for some RG flow connecting these different fixed points.
The knowledge of the partition function with fundamentals and adjoints made also possible the study of some higgsing flow, and even in that case we verified
 the validity of the $F$-theorem.
Then, we moved to the gauge theories with two gauge groups, built from the ABJ(M) models. In this case we have a large spectrum of fixed points and RG flows
 among them where we can compare our results with the perturbative series and look for further evidence of the $F$-theorem. Our results point towards a positive
 answer for its existence.

Many interesting questions arise from 
the study of the localized partition function of $3d$ $\mathcal{N}=2$ SUSY gauge theories.
First it is necessary to understand the sign of $\partial^2_a \mathcal{|Z|}$. In all the examples 
computed till now this derivative is always positive. It induces the conjecture that the free energy counts massless degrees of freedom, decreasing at the endpoints of an RG flow.

A deeper investigation is then required for the quivers with  chiral like field content
studied in \cite{Martelli:2008si,Hanany:2008cd,Ueda:2008hx,Hanany:2008fj,
Franco:2008um,Hanany:2008gx,Franco:2009sp,Davey:2009sr,Davey:2009et,
Aganagic:2009zk}. 
In  \cite{Jafferis:2011zi} it was observed that at large $N$ the free energy do not scale as $N^{3/2}$ as their conjectured gravity duals.\footnote{ This scaling
 was first computed in \cite{Drukker:2010nc,Herzog:2010hf} for $\mathcal{N}\geq 3$ SUSY.}
On the contrary chiral theories with extra flavor \cite{Benini:2009qs,Jafferis:2009th}
have the expected entropy scaling.
 It would be interesting to study, where it is possible, the weakly coupled limit and
  see if these models are counterexamples of the $F$-theorem as well.
Another possible source of  counterexamples is represented by the appearance of accidental symmetries in the IR.  For example in 
theories with adjoint matter interacting via a high degree polynomial superpotential
many accidental symmetries in the IR are generated.  Without modifying the free energy these accidental symmetries spoil the $F$-theorem. Thus, it is
 necessary to modify  the partition function or the free energy. In  \cite{Niarchos:2011sn} a preliminary discussion about this topic was given,
starting from the numerical computation of the free energy in theories with accidental symmetries.
Anyway it would be interesting to have some analytic example to understand the functional behavior of the $R$ charge
in these strongly coupled theories, along the lines of \cite{Minwalla:2011ma}.

In this paper we found only a possible source of violation of the theorem in the flavored ABJ theories. Indeed
for some accidental values of $N_f$ and $k$ we observed that $\Delta F$ vanishes for the $SU(2) \times SU(2)$ gauge group
at four loop. Computations for general $N$ and four higher loop become then necessary to check the validity of the theorem.

Another interesting topic is related to the localization of the partition function in $\mathcal{N}=2$ four dimensional SUSY 
gauge theories \cite{Pestun:2007rz} and its possible relation with the computation of the conformal anomaly given in \cite{Shapere:2008zf}.
 This possible agreement would suggest another connection between the four dimensional a-theorem and the F-theorem.

A stronger version of the theorem may be formulated by the knowledge of the free energy out of the fixed point.
In analogy with the four dimensional case a possibility can be the insertion of some Lagrange multiplier, multiplying the 
constraints imposed by the beta function. Out of the fixed point we expect $\mathcal{Z}\equiv\mathcal{Z}(\lambda,\Delta)$
where $\lambda$ are the coupling constant of the theory.  If these coupling constant are interpreted as Lagrange multiplier 
we expect that
\begin{equation}
\mathcal{Z}(\lambda,\Delta) = e^{-F(\Delta)+\lambda_i \beta_{\lambda_i}}
\end{equation}
because the free energy has a role similar to the four dimensional conformal anomaly. 
We leave the verification of this statement along the lines of \cite{Kutasov:2003ux} for future works.
An interesting connection between this possible extension of the $F$-extremization
and gauge gravity duality may be provided by the recent results of \cite{Myers:2010xs,Myers:2010tj}.
In that case the holographic studies lead
the authors to identify the quantity that flows monotonically using a calculation of
entanglement entropy  in the fixed point CFTs. 
in four dimensions the entanglement entropy
corresponds to the conformal anomaly at the fixed point.
More generally in \cite{Casini:2011kv} a connection between the entanglement entropy
and the partition function on $S^d$ was found. It would be interesting to understand
if the connection can be extended out of the fixed point, and to check 
if the Lagrange multiplier are promising for this extension.

Finally it would be important to check the agreement of the partition function among Seiberg dual phases.
The notion of four dimensional Seiberg duality was extended to three dimensional YM theories
in \cite{Aharony:1997bx,Karch:1997ux,Aharony:1997gp}  and for theories with CS terms in \cite{Giveon:2008zn} and extended in \cite{Niarchos:2008jb} to the case of adjoint matter.
Some preliminary results for theories with $\mathcal{N} \geq3 $ supersymmetry and for $\mathcal{N}=2$ theories with one gauge group
appeared in \cite{Kapustin:2010xq,Willett:2011gp,Kapustin:2011gh}.
In the case of multiple gauge groups some extra rule is necessary (see \cite{Aharony:2008gk} for the case with two groups and
\cite{Amariti:2009rb} for an extension to N groups).
Moreover the recent connection among the four dimensional superconformal index and the three dimensional partition function 
 \cite{Dolan:2011rp,Gadde:2011ia,Imamura:2011uw}, may be useful to generate and check new type of Seiberg dualities, as observed in \cite{Dolan:2011rp}.

\section*{Acknowledgments}

We are grateful to Ken Intriligator for discussions.
A.A. is supported by UCSD grant DOE-FG03-97ER40546. The work of M.S. is supported in part by the FWO - Vlaanderen,
Project No. G.0651.11, and in part by the Federal Office for Scientific,
Technical and Cultural Affairs through the ``Interuniversity Attraction
Poles Programme -- Belgian Science Policy'' P6/11-P.

\appendix
\section{Useful formulas and conventions} \label{A}

For $SU(N)$ we use the $N\times N$ hermitian matrix generators $T^a$ ($a=1,\ldots, N^2-1$) 
normalized as $\Tr T^A T^B = \delta^{AB}$. The structure constants are defined by $\left[ T^A, T^B \right]=i f^{ABC}\, T^C$. Accordingly, the second
 Casimir in the fundamental and adjoint
representations reads
\begin{eqnarray}
 C(N) &=& \frac{N^2-1}{N} \nonumber \\
 C(G) &=& 2 N \nonumber
\end{eqnarray}

\vskip 10pt
The Feynman diagrams are computed by performing the integrals in momentum space and using dimensional
regularization ($d=3 -2\epsilon$). The relevant one-loop integrals are
\begin{eqnarray}
  \intk{k} \frac{1}{k^2 (k-p)^2} &=& \frac{1}{8} \frac{1}{|p|} \equiv
  B_0(p) \label{1integral} \\
  \intk{k} \frac{k_{\a\b}}{k^2 (k-p)^2} &=& \frac12 \, p_{\a\b} \, B_0(p)
\end{eqnarray}
while at the two-loop level we only need
\begin{eqnarray}
F(p) \equiv   \intkk{k}{q} \frac{1}{k^2\, q^2\, (p-k-q)^2} = \frac{\Gamma(\epsilon)}{64\pi^2} \sim \frac{1}{64\pi^2}
\, \frac{1}{\epsilon} 
\label{integral}
\end{eqnarray}

Given the above expressions, the anomalous dimensions (\ref{eq:pert1}) can be written in terms of gauge group invariants as
\begin{equation}
\begin{split}
 \gamma_q &= - \frac{1}{2 k^2} \, C(N) \left[ C(N) + \frac{C(G)}{2} (M-1) + N_f \right] \\
 \gamma_\phi &= - \frac{1}{2 k^2} \, C(G) \left[ \frac{C(G)}{2} (M+1) + N_f \right]
\end{split}
\end{equation}
or more compactly, for a field in the representation $r$ of the gauge group
\begin{equation}
 \gamma_r = - \frac{1}{2 k^2} \, C(r) \left[ C(r) + \frac{C(G)}{2} (M-1) + N_f \right]
\end{equation}
Finally, the superpotential contributions (\ref{eq:deltagamma}) are written as
\begin{equation}
 \delta\gamma_r = \frac{|\a_2|^2}{32 \pi^2} \, {\cal M}_r \, \frac{|G|}{|r|} \left( 2 C(N) - \frac{1}{2} C(G) \right)
 \label{eq:deltaWgamma}
\end{equation}
where ${\cal M}_r$ is the number of tree-level couplings between the field in representation $r$ and the other fields in the theory (for instance, the
fundamentals are coupled to ${\cal M}_q=M$ adjoints, while the adjoints are coupled to ${\cal M}_\phi=2 N_f$ fundamentals).\footnote{It would be
interesting to check if a formula similar to (\ref{eq:deltaWgamma}) also holds for other representations.} Similar formulas can be derived for the two gauge
groups case.

\section{Integrals and Series for the computation of the partition function} \label{B}
In this appendix we list the integrals and the series that we used to perform the computation
for the $SU(N)_k$ theory.
The generic integral is 
\begin{equation}
I = \int 
\prod_{i}\text{d}u_i e^{i k \pi \Tr u^2}f(\{u_i\}) \Delta^2(2 \pi u) 
\end{equation}
If $f(\{u_i\})=1$ we obtain 
\begin{equation}
 I_1 = \sqrt{(-1)^N} e^{-\frac{1}{4} i N^2 \pi } k^{-\frac{N^2}{2}} 
\left(2 \pi \right)^{\frac{(N-1) N}{2} } \text{G}_2 (N+2) 
\end{equation}
where G$_2$ is the Barnes function defined by G$_2(z+1)=\Gamma(z) $G$_2(z)$,
and G$_2(1)=1$.
For different values of the function $f(\{u_i\})$ the integrals are computed 
from the recursive relation of the orthogonal polynomials.
The integrals that are necessary 
for our computation are listed below.
\begin{equation}
\begin{array}{|l|c|c|}
\hline
f(\{u_i\})&&I/I_1\\
\hline
\sum u_i^2&
&
\frac{i N^2}{2 \pi k}
\\
\sum u_i u_j &(i<j)
&
-\frac{i N(N-1)}{4 \pi k}
\\
\sum_{i} u_i^4&&
-\frac{N(2 N^2+1)}{4 \pi^2 k^2}
\\
\sum u_i^2 u_j^2&(i<j)
&
-\frac{N(N-1)(N^2-N+1)}{8 \pi^2 k^2}
\\
\sum u_i^2 u_j u_k & (j<k)
&
\frac{N(N-1)(N-2)^2}{8 \pi^2 k^2}
\\
\sum u_i^3 u_j
&&
\frac{N(N-1)(2 N-1)}{4 \pi^2 k^2}
\\
\sum u_i u_j u_k u_l &(i<j<k<l)
&
-\frac{N(N-1)(N-2)(N-3)}{32 \pi^2 k^2}\\
\hline
\end{array}
\end{equation}
Moreover we have to compute some series
coming from the Fourier decomposition of the $\delta$-function.
They are
\begin{eqnarray}
\sum_{m=-\infty}^{\infty} e^{-\frac{m^2}{ 2 \alpha}} &=& \sqrt{2 \pi \alpha}\equiv S_0
 \nonumber \\
 \sum_{m=-\infty}^{\infty} \frac{m^2}{k^2}  e^{-\frac{i N m^2}{4 k \pi}}&=&-\frac{ 4 \pi i}{N} \partial_{k} S_0
\nonumber \\
\sum_{m=-\infty}^{\infty} \frac{m^4}{k^4}  e^{-\frac{i N m^2}{4 k \pi}}&=& -\frac{16 \pi^2}{N}\left(\partial_k^2+\frac{2}{k}\partial_k\right)S_0
\end{eqnarray}

\bibliographystyle{ieeetr}
\bibliography{CFfth}

\end{document}